\documentclass{INTERSPEECH2023}

\interspeechcameraready

\usepackage{algorithm}
\usepackage{amssymb,amsfonts}
\usepackage{algpseudocode}
\usepackage{textcomp}
\usepackage{color}
\usepackage{multirow}
\usepackage{subfig}
\usepackage{float}

\hypersetup{
	colorlinks=true,
	linkcolor=red,
	filecolor=blue,      
	urlcolor=red,
	citecolor=green,
}

\title{Multi-Scale Attention for Audio Question Answering}
\name{Guangyao Li$^1$, Yixin Xu$^1$, Di Hu$^{1,2,*}$}
\address{
  $^1$Gaoling School of Artificial Intelligence, Renmin Uniiversity of China, Beijing \\
  $^2$Beijing Key Laboratory of Big Data Management and Analysis Methods, Beijing
  }
\email{\{guangyaoli, xu\_yixin, dihu\}@ruc.edu.cn}

\begin{document}
\vspace{-1.5em}

\maketitle
 
\begin{abstract}
Audio question answering (AQA), acting as a widely used proxy task to explore scene understanding, has got more attention.
The AQA is challenging for it requires comprehensive temporal reasoning from different scales' events of an audio scene. 
However, existing methods mostly extend the structures of visual question answering task to audio ones in a simple pattern but may not perform well when perceiving a fine-grained audio scene. 
To this end, we present a \textbf{M}ulti-scale \textbf{W}indow \textbf{A}ttention \textbf{F}usion \textbf{M}odel (\textbf{MWAFM}) consisting of an \textit{asynchronous hybrid attention module} and a \textit{multi-scale window attention module}.
% The former is designed to aggregate unimodal and cross-modal temporal contexts, the latter captures sound events with various lengths and their temporal contextual dependencies for better understanding.
The former is designed to aggregate unimodal and cross-modal temporal contexts, while the latter captures sound events of varying lengths and their temporal dependencies for a more comprehensive understanding. 
Extensive experiments are conducted to demonstrate that the proposed MWAFM can effectively explore temporal information to facilitate AQA in the fine-grained scene.
% Code is available in \href{http://gewu-lab.github.io/MWAFM/}{http://gewu-lab.github.io/MWAFM/}
\footnote{Code: \href{https://github.com/GeWu-Lab/MWAFM}{https://github.com/GeWu-Lab/MWAFM} }

% Audio Question Answering (AQA) has become increasingly popular as a proxy task for scene understanding, but it remains challenging due to the need for comprehensive temporal reasoning across multiple scales of events in an audio scene. 
% While many existing methods have simply extended visual question answering structures to audio tasks, they may not perform well in perceiving fine-grained audio scenes. 
% To address this, we propose the \textbf{M}ulti-scale \textbf{W}indow \textbf{A}ttention \textbf{F}usion \textbf{M}odel (\textbf{MWAFM}), which incorporates an asynchronous hybrid attention module and a multi-scale window attention module. 
% The former aggregates unimodal and cross-modal temporal contexts, while the latter captures sound events of varying lengths and their temporal dependencies for a more comprehensive understanding. 
% Our extensive experiments demonstrate that the MWAFM effectively explores temporal information to facilitate AQA in fine-grained scenes.
\end{abstract}

\begin{figure*}[ht]
     \centering
     \includegraphics[width=0.95\textwidth]{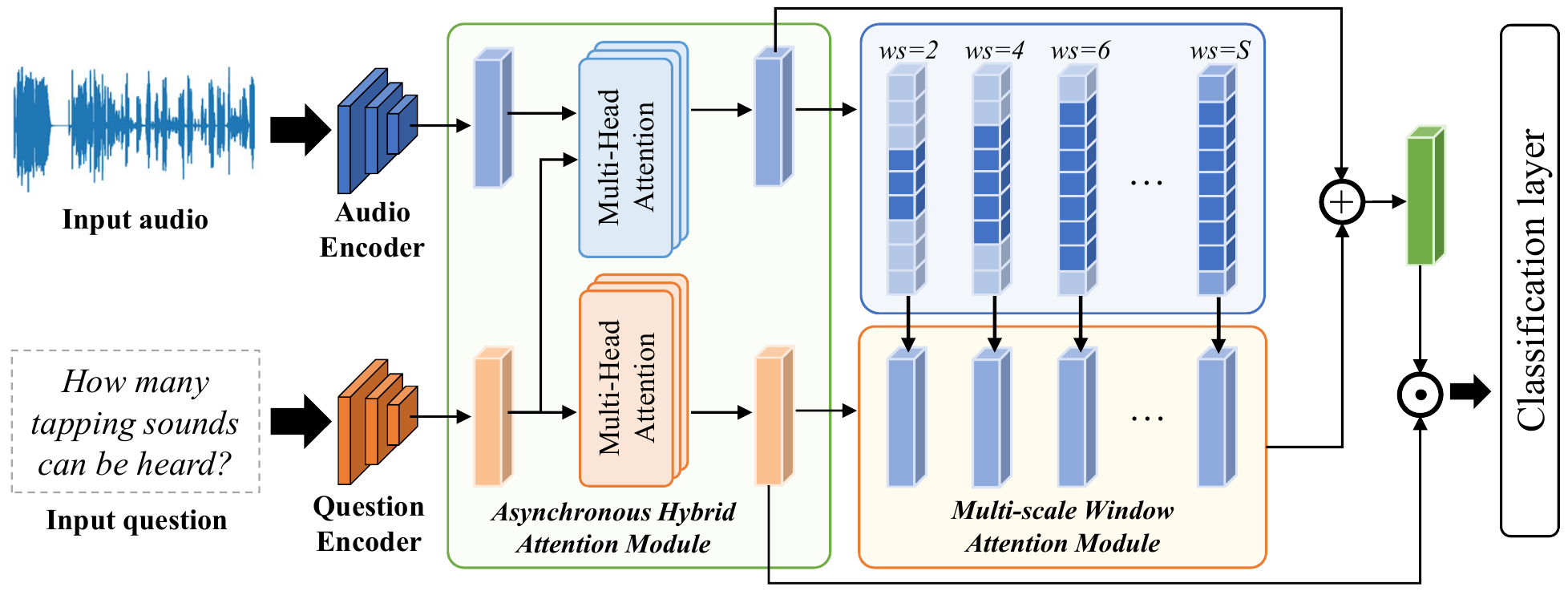}
     \caption{The pipeline of our proposed  \textbf{M}ulti-scale \textbf{W}indow \textbf{A}ttention \textbf{F}usion \textbf{M}odel (\textbf{MWAFM}). The model takes pre-trained CNNs to extract audio features and uses a Fasttext pre-trained word vectors to obtain question features. Firstly, we aggregate unimodal and cross-modal temporal contexts. Then we highlight audio features of key timestamps for temporal multi-scales information through question queries. Finally, multimodal fusion is exploited to integrate audio and question information for predicting the answer to the input question. (\textit{\textbf{ws}} indicate \textbf{w}indow \textbf{s}ize).}
     \label{fig:framework}
     \vspace{-0.75em}
\end{figure*}

\section{Introduction}

We are surrounded by a complex mixture of audio signals in daily life, and our auditory perception system unconsciously focuses on sound sources of interest~\cite{mesaros2021sound}.
Imagine sitting on your sofa at home with your eyes closed, and you can hear and recognize a succession of sounds: baby crying, family members speaking, their footsteps, etc.
Understanding all these sounds and interpreting the perceived scene as a domestic scene comes naturally to humans but is still challenging for machines.
Hence, making machines leverage audio information, especially authentic sounds in natural scenes, to achieve considerable audio scene perception and understanding ability as humans is an interesting and valuable topic. 

Recently, we have seen significant progress in sound event detection~\cite{cakir2017convolutional, mnasri2022anomalous},  speech recognition~\cite{baevski2021unsupervised,li2022recent} and enhancement~\cite{michelsanti2021overview}, music transcription~\cite{benetos2018automatic}, audio source separation~\cite{makino2018audio}, and audio question answering~\cite{fayek2020temporal,lipping2022clotho} toward audio scene understanding.
Among them, question answering has been widely used as a proxy task to explore scene understanding along with getting more and more attention.
Inspired by CLEVR~\cite{johnson2017clevr}, some researchers proposed a program-generated dataset containing fixed-length audio sequences of different notes to explore the AQA task~\cite{abdelnour2018clear}. 
Following this trajectory, the DAQA~\cite{fayek2020temporal} and NAAQA~\cite{abdelnour2022naaqa} are proposed to extend the questions to more acoustically realistic situations and have achieved outstanding achievements.
However, the generated data usually lack diversity and challenges in the natural scene.
To this end, some others tend to focus on answering questions with spoken~\cite{zhang2017speech} and voice~\cite{chuklin2019using} information and answering questions about videos might require joint reasoning about visual and audio cues.
For the audio scenes above main consist of human speech, to explore more natural sound scene understanding, Clotho-AQA~\cite{lipping2022clotho} is proposed that contains audio files of day-to-day sounds occurring in the environment, such as birds, etc., while avoiding human speech. 
Meanwhile, MUSIC-AVQA~\cite{Li_2022_CVPR} also includes many natural sounds and includes the AQA sub-task. 
These datasets provide a good testbed for the AQA task and attract researchers' attention. 
% However, existing solutions of AQA tasks on the dataset above are mostly similar to the methods of VQA when dealing with the temporal reasoning problems. 
To our knowledge, current research on AQA is mostly presented in the form of new datasets and baseline networks (e.g., DAQA, Clotho-AQA), 
% without taking into account the characteristics of sound scenes, such as event duration and long-term dependence.
but ignores the characteristics of sound scenarios such as event duration and long-term dependence.
Additionly, reasoning exploration in VQA tasks focuses on spatial or spatio-temporal (e.g., Audio-MUSIC-AQA) problems, and its spatial information plays an indispensable role in model, which is not suitable for directly transferring reasoning network in VQA to AQA tasks.
Hence, how to explore temporal reasoning problems in AQA tasks effectively, especially capturing sound events at different scales and their correlation, is challenging but significant for fine-grained sound scene understanding.

In this work, we propose a  \textbf{M}ulti-scale \textbf{W}indow \textbf{A}ttention \textbf{F}usion \textbf{M}odel (\textbf{MWAFM}) on Clotho-AQA and Audio-MUSIC-AVQA dataset.
Concretely, we apply a new asynchronous hybrid attention module (AHAM) to leverage unimodal and cross-modal temporal contexts. 
Then, given that multi-length sound event, whose temporal context is crucial for understanding complex scenes, a develop multi-scale window attention module (MWAM) to capture sound events of different scales. 
Furthermore, since the audio event changes over time dynamically, the model uses question features as queries to attend crucial temporal segments for encoding question-aware audio embedding effectively. 
As an open-ended problem, the correct answers to questions can be predicted by choosing words from a predefined answer vocabulary. 
% To validate the effectiveness of the proposed model, 
To validate the proposed model, 
we conducted a large number of experiments, including comparison with mainstream QA methods, various ablation study module et al,. 
Sufficient experimental results demonstrate the effectiveness and generalization of the proposed MWAFM.
% Our results indicate that our model outperforms recent AQA and other QA-related approaches.

% The main contributions of this paper include follows:
The main contributions include:
\textbf{1)} A novel asynchronous hybrid attention module to leverage unimodal and cross-modal temporal contexts.
\textbf{2)} An effective attentive multi-scale window attention module captures sound events of various lengths and their temporal contextual dependencies.
\textbf{3)} The proposed MWAFM provides a powerful ability for AQA tasks, which achieves superior performance on two public benchmarks and demonstrates the model's effectiveness and generalization.

\section{Method}
To facilitate audio question answering task effectively, we proposed \textbf{M}ulti-scale \textbf{W}indow \textbf{A}ttention \textbf{F}usion \textbf{M}odel (\textbf{MWAFM}), which is composed of an asynchronous hybrid attention module (AHAM), and a multi-scale window attention module (MWAM).
The architecture of \textbf{MWAFM} is illustrated in Figure.~\ref{fig:framework}.

% \vspace{-0.4em}
\subsection{Representations for Different Modalities}
\textbf{Audio Representation}. 
Given an input audio sequence, we first divide it into $T$ non-overlapping audio segments $A=\{a_1, a_2, ...,a_T\}$, where each segment is 1$s$ long, and $T$ is the timestamp. We encode each audio segment $A_t$ into a feature vector $f^t_a$ using a pre-trained VGGish model~\cite{gemmeke2017audio}, which is a VGG-like 2D CNN network, employing over transformed audio spectrograms. The audio representation is extracted offline and the model is not fine-tuned.

\textbf{Question Representation}. 
For an asked question $Q = \{q_n\}^N_{n=1}$ input, 
the Fasttext pre-trained model~\cite{mikolov2018advances} is used to process projected word vectors.
If the input question $Q$ has $N$ words, the word embedding shape using Fasttext is $N\times300$.
And the word embeddings are passed through a linear layers into a feature vector $f_q$.
% and encode the question into a feature vector $f_q$ using the last hidden state. The question encoder is trained from the scratch.

%% Unidirectional部分
\subsection{Asynchronous Hybrid Attention Module}
Natural audios tend to contain continuous and repetitive rather than isolated sound event. Especially, sound events usually redundantly recur many times.
To capture multimodal temporal contexts, we design a new temporal mechanism: Asynchronous Hybrid Attention module (AHAM), which uses two unimodal encoders and a multimodal encoder to learn which snippets for each audio adaptively.

\textbf{Unimodal Encoder}. For each of the question and audio input representations $\{f_q^n\}_{n=1}^N, \{f_a^t\}_{t=1}^T$, we first apply the layer normalization and feed them into the corresponding unimodal encoder, in which we use self-attention module to update $f_q^n$ and $f_a^t$, respectively. 
An asynchronous hybrid attention function $H$ in AHAM will be learned from audio and question features: $\{q_i\}_{i=1}^N, \{a_i\}_{i=1}^T$ to update $f_q^n$ and $f_a^t$, respectively. 
The updated audio and question feature $\hat{f}_a^t$, $\hat{f}_q^n$ can be obtained with same computation:
\begin{align}
\hat{f}_m^l = f_m^l + \sum_{l=1}^{\intercal}w_l f_m^l
=f_m^l + {\sigma}(\frac{f_m^l\textbf{\textit{f}}_m^{\intercal}}{\sqrt{d}})\textbf{\textit{f}}_m\enspace,
\end{align}
where $m$ is the question or audio modality, and $l$ is their corresponding length;
$f_a^t = [f_a^1;...;f_a^T]$ and $f_q^n = [f_q^1;...;f_q^N]$; 
% $H_{sa}$ is self-attention function; 
skip-connections can help preserve the identity information from the input sequences. 
% The two unimodal encoder are formulated with the same computation mechanism and they are defined as:
% where the formula is audio as an example, the same is true for question.
The $\sigma$ is softmax function; the $d$ is a scaling factor with the same size as the feature dimension and $(\cdot)^{\intercal}$ denotes the transpose operator.

\textbf{Multimodal Encoder}. 
To highlight the audio key timestamps closely associated with the question, we utilize the asynchronous cross-modal attention, designed to attend critical temporal segments among the changing audio scenes and capture question-aware audio embeddings.
Concretely, given $\{f_q^n\}_{n=1}^N$ and audio features $\{f_a^t\}_{t=1}^T$ from input embedding, the multimodal encoder will learn to aggregate question-aware audio features.
The module will produce a more robust audio feature representation $F_a^t$:
\begin{align}
F_a^t = f_a^t + \sum_{t=1}^{\intercal}w_t^af_a^t= f_a^t + {\sigma}(\frac{f_q\textbf{\textit{f}}_a^{\intercal}}{\sqrt{d}})\textbf{\textit{f}}_a\enspace.
\end{align}
% where $H_{ca}$ is cross-attention function.
The AHAM will assign large weights to audio snippets which are more relevant to the question.
Then the obtained audio and question features $F_a, F_q=\hat{f}_q$ are input to the MWAM.

%% Paragraph 02: multi-scale window-size 部分
\subsection{Multi-scale Window Attention Module}
\begin{figure}[tp]
     \centering
     \includegraphics[width=0.475\textwidth]{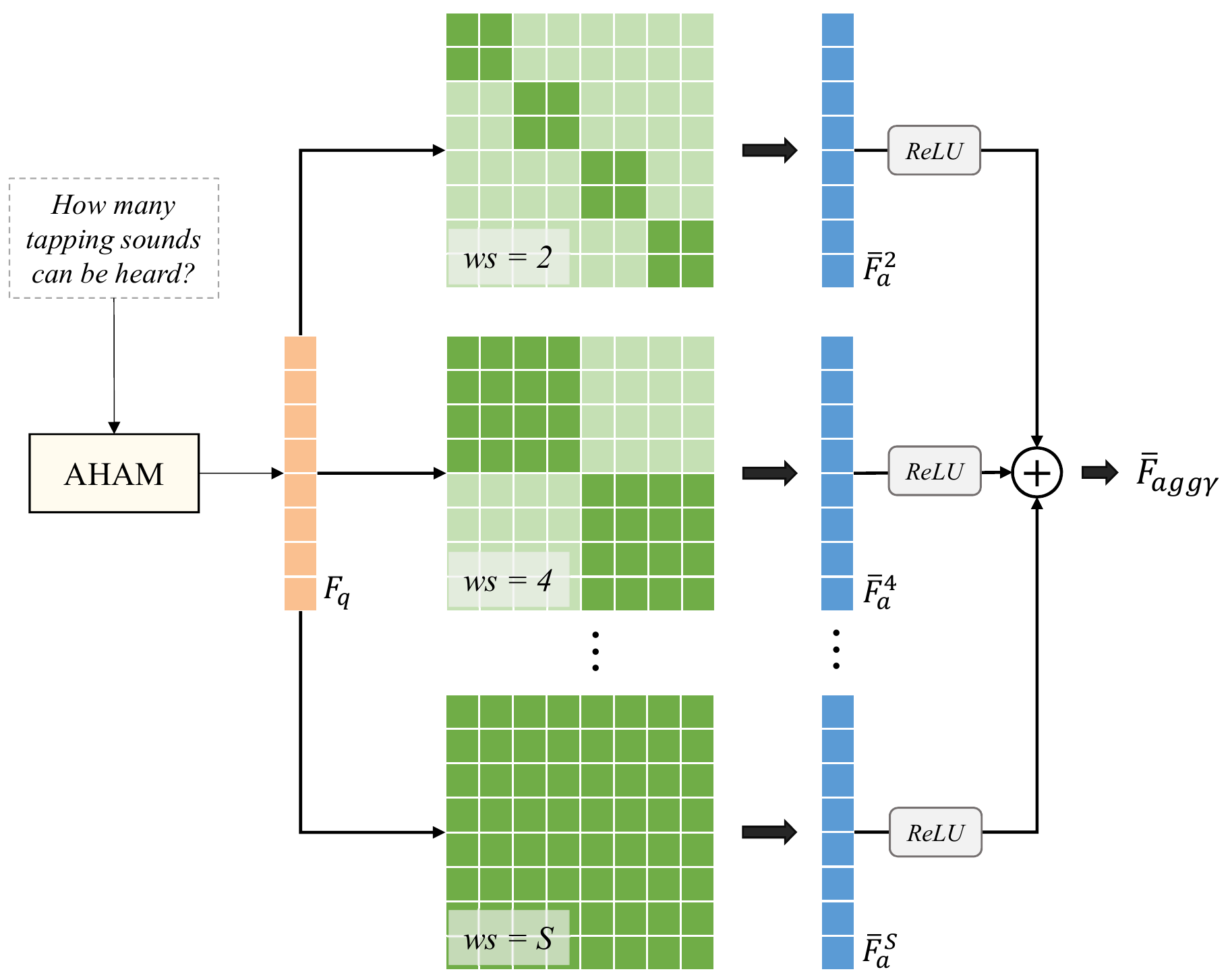}
     % \vspace{-0.5em}
     \caption{Multi-scale Window Attention Module (\textbf{MWAM}). The Module will aggregate multi-scale question-aware audio features (\protect \includegraphics[scale=0.45]{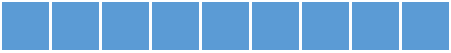}), which is obtained through question queries on the output from sliding window of different size.}
     \label{fig:multi-scale}
     \vspace{-1.5em}
\end{figure}

To capture the intrinsic property that different events have various duration, the multi-scale window attention with different window size is designed, such a stacked shifted window transformer with window size increasing with layers becoming deeper.
Given the importance of local context, our attention pattern employs multi-size window attention surrounding each token.
Given a fixed window size $S$, each token attends to $S/2$ on each side (The dark part in the green area in Figure.~\ref{fig:multi-scale}).
For different scales of $S$, we compute the self-attention score of the $S$ window size on audio, where $Q\text{-}value, K\text{-}value, V\text{-}value$ are the corresponding values in the sliding window. The audio feature expression $\hat{F}_a^i$ at different scales by:
\begin{align}
\hat{F}_a^i = \mathit{MHAttn}(F_a^i, F_a^i, F_a^i), i = 2, 4, 6, ..., S,
\end{align}
where $S$ is the size of the sliding window. 
Then, the model will learn to aggregate question-aware audio features on different scales $\hat{F}_a^i$ by \textit{Equation (2)}. And the output feature $\bar{F}_a^i$ can be computed as: 
\begin{align}
\label{ws}
\bar{F}_a^i = \mathit{MHAttn}(F_q, F_a^i, F_a^i), i = 2, 4, 6, ..., S.
\end{align}

We aggregate $\bar{F}_a^i$ and $F_a$ to better capture multi-scale sound events under the guidance of question by:
\begin{align}
\bar{F}_{aggr} = Norm(\sum_i^S(ReLU(\bar{F}_a^i) + ReLU(F_a)))\enspace,
\end{align}
where $\bar{F}_{aggr}$ is audio feature through MWAM and AHAM.
The illustration of \textit{Equation (5)} can be seen in the blue part at the bottom of Figure~\ref{fig:multi-scale}. Among them, a linear projection and a dropout layer are operated before \textit{ReLU}.
Hence, the quesiton queried audio contextual embeddings are more capable for predicting correct answers.

\subsection{Feature Fusion and Answer Prediction}
Different modalities can contribute to correctly answer questions. 
To combine the features: $\bar{F}_{aggr}$ and $F_q$, we introduce a simple multimodal fusion network. 
Specifically, we integrate audio and question features with employing an element-wise multiplication operation by $e = \bar{F}_{aggr}\circ F_q$. 
% Concretely, we can formulate the fusion function as: $e = \hat{F_a}\circ F_q$. 

To achieve AQA task, we predict the answer for a given question based on the joint embedding $e$. 
It can be formulated as an open-ended task, which aims to choose one correct word as the answer from a pre-defined answer vocabulary. 
We utilize a linear layer and softmax function to output a probability $p \in \mathcal{R}^{C}$ for candidate answers.
With the predicted probability vector and the corresponding ground-truth label $y$, we can optimize our network using a cross-entropy loss:
\begin{align}
\mathcal{L}_{qa} = -\sum_{c=1}^{C} y_clog(p_c).
\end{align}
During testing, we can select the predicted answer by $\hat{c} = \arg \text{max}_c(p)$.

\section{Experiments and Analysis}
% \vspace{-0.5em}
\label{sec:pagestyle}
\subsection{Datasets}
To benchmark the performance of \textbf{MWFAM}, we conduct experiments on two large-scale datasets, including Clotho-AQA and Audio-MUSIC-AVQA.
\textbf{Clotho-AQA}, an audio question answering dataset consisting of 1,991 audio files selected from the Clotho~\cite{drossos2020clotho} each between 15 to 30 seconds in duration.
The training, validation and test splits of Clotho-AQA contain 1,174, 344, and 473 audio files and 828, 512 and 801 unique answers without \textit{yes/no}, respectively.
\textbf{Audio-MUSIC-AVQA} is a subset of MUSIC-AVQA~\cite{Li_2022_CVPR} dataset that only contains audio-related question-answer pairs and the corresponding audio.
We split the dataset into training, validation, and testing sets for training and evaluation with 5,633, 806, and 1,611 QA pairs, respectively.

\subsection{Evaluation}
We follow the standard evaluation protocol of each dataset.
We use answer prediction accuracy as the metric and evaluate model performance on answering different types of questions. 
It should be noted that we use \textit{Top}-1, \textit{Top}-5, and \textit{Top}-10 accuracy metrics to evaluate the performance of the Clotho-AQA dataset because the number of unique answer classes is high (828).
But we only use \textit{Top}-1 accuracy as a metric on the Audio-MUSIC-AVQA dataset because its answer vocabulary consists of 42 possible answers to different questions.
For training, we use one single model to handle all questions without training separated models for each type.

\begin{table}[t]
\begin{center}
\caption{AQA results of different methods on the test set of Chloto-AQA. The top-1 results are highlighted.}
\label{table01_compare}
\vspace{-0.5em}
\scalebox{0.95}{
\begin{tabular}{c|ccc}
\hline
\textbf{Method}        & \textbf{Top-1 Acc} & \textbf{Top-5 Acc} & \textbf{Top-10 Acc} \\ \hline
GRU~\cite{antol2015vqa}                   & 09.21  & 26.02   & 36.87    \\
BiLSTM~\cite{hori2017attention}           & 14.39  & 35.90   & 47.00    \\
HME~\cite{fan2019heterogeneous}           & 12.79  & 32.85   & 42.54         \\
PSAC~\cite{li2019beyond}	              & 14.29  & 35.80   & 46.32    \\
LongFormer~\cite{beltagy2020longformer}   & 15.98  & 36.72   & 47.67    \\
FCNLSTM~\cite{fayek2020temporal}          & 20.16  & 44.43   & 55.57    \\
CONVLSTM~\cite{fayek2020temporal}         & 18.99  & 40.70   & 49.27    \\
ST-AVQA~\cite{Li_2022_CVPR}               & 17.20  & 40.16   & 50.73    \\ 
AquaNet~\cite{lipping2022clotho}          & 14.78  & 36.19   & 46.71    \\
\hline
\textbf{MWAFM}  & \textbf{22.24} & \textbf{46.56} & \textbf{57.75}\\
\hline
\end{tabular}
}

\end{center}
\vspace{-1.5em}
\end{table}

\subsection{Implementation  Details}
% \vspace{-0.25em}
The audio stream is divided into non-overlapping 1s segments, and the sampling rate of sounds is 16 k$Hz$.
We use a linear layer for each 1s-long audio segment to process the extracted 128-$D$ VGGish feature into a 512-$D$ feature vector. 
The audio is fixed to 24 seconds through the \textit{interpolate} operation, and the question length is fixed to 20.
Batch-size and number of epochs are 64 and 50, respectively. 
The initial learning rate is $1e\text{-}4$, and our network is trained with the Adam optimizer.
All the experiments are conducted on NVIDIA-V100 GPUs.

% \vspace{-0.25em}
\subsection{Results and analysis}
% \vspace{-0.25em}
In this subsection, we evaluate the proposed MWAFM on Clotho-AQA and Audio-MUSIC-AVQA to present the model performance for the AQA task.

\textbf{Comparison with QA Models.}
To validate our MWAFM on the Clotho-AQA dataset, we compare it with recent QA methods: 
GRU~\cite{antol2015vqa}, BiLSTM~\cite{hori2017attention}, AquaNet~\cite{lipping2022clotho},
% HCAttn\cite{lu2016hierarchical}, MCAN~\cite{yu2019deep}, 
HME~\cite{fan2019heterogeneous}, LongFromer~\cite{beltagy2020longformer}, 
FCNLSTM~\cite{fayek2020temporal}, CONVLSTM~\cite{fayek2020temporal},
ST-AVQA~\cite{Li_2022_CVPR}.
Specifically, FCNLSTM is applied to temporal reasoning of sound events, but it is difficult to capture multi-lengths events because of its only modeling of global information. AquaNet is designed for AQA tasks, but it simply migrates from VQA methods without reasoning capabilities. Although the ST-AVQA method is dedicated to real-world scene understanding and reasoning, it combines audio and visual modalities to improve the perception ability of the model. For the scene containing only audio modality, it is difficult for the temporal module of the ST-AVQA to carry out fine-grained understanding.
Tabel.~\ref{table01_compare} shows results of those comparable methods, where we only use the \textit{Temporal Grounding Module} in the ST-AVQA method.
The results demonstrate that our model achieves considerable improvement on most QA methods.
It is worth noting that our method achieves excellent performance on which metrics are \textit{Top}-1, \textit{Top}-5, and \textit{Top}-10, indicating the effectiveness of the proposed MWAFM.

% \textcolor{blue}{
% The official Cloto-AQA dataset provides two versions, one containing \textit{yes} and \textit{no} and the other not. Considering that the former version is not cleaned and biased, the Cloto-AQA version used in the experiment not contain \textit{yes} and \textit{no}.
% }

\begin{table}[t]
\begin{center}
\caption{Ablation study on different window-size and the proposed modules. We observe that leveraging UHAM and MWAM can boost AQA task. (\textit{\textbf{ws}} indicate \textbf{w}indow \textbf{s}ize).}
\label{ablation}
\vspace{-0.65em}
\scalebox{0.99}{
\begin{tabular}{c|c|ccc}
\hline
Method         & \textbf{ws}      & \textit{Top}-1     & \textit{Top}-5     & \textit{Top}-10  \\ 
\hline
w/ ws-attn     & 2              & 21.07     & 45.10     & 55.64   \\
w/ ws-attn     & 4              & 21.27     & 44.57     & 55.45   \\
w/ ws-attn     & 6              & 21.42     & 44.90     & 56.07   \\
w/ ws-attn     & 12             & 21.32     & 45.17     & 56.06   \\
\hline
w/o un-queried & 2,4,6,12       & 20.30	    & 43.51	    & 54.07   \\
w/ bi-queried  & 2,4,6,12       & 21.17     & 43.22     & 55.57   \\
w/o AHAM       & 2,4,6,12       & 13.32     & 36.14     & 47.24   \\
w/o MWAM       & -              & 21.75     & 44.48     & 55.23   \\
\hline
MWAFM$^{\prime}$          & 12,12,12,12    & 20.35	    & 43.17	    & 54.89   \\
MWAFM*         & ——    & 21.22	    & 46.08	    & 56.44   \\
\textbf{MWAFM} & 2,4,6,12   & \textbf{22.24} & \textbf{46.56} & \textbf{57.75} \\
\hline
\end{tabular}
}
\end{center}
\vspace{-1.25em}
\end{table}

\textbf{Ablation study.} 
% Table.2 shows the results of the ablation study. The model, which only use single window without aggregation, achieves accuracy not perform well on all three metrics.
% State-of-the-art performance is achieved when the model aggregates features at multiple scales.
% It is proved that the multi-scale fusion strategy model can improve the performance of AQA task.
% Table 2 shows the results of the ablation study.
% Table 2 shows the effect of AHAM and MWAM on the proposed model performance.
% When the model uses only one window without aggregation, the results are poor on all three evaluation metrics.
% Excellent performance can be obtained only when the model aggregates feature at multiple scales.
% It demonstrated that the multi-scale fusion policy model could well capture sound events and their dependencies at different scales.
% Table 2 shows the effects of AHAM and MWAM and their components on model performance.
% Obviously, we can see that both AHAM and MWAM can improve the performance of the model.
% Specifically, the Multimodal Encoder in AHAM only uses the question to query the audio better than they directly query each other, indicating that the semantics of the sentence needs to be associated with a certain segment of the audio in the QA question.
% Furthermore, the model aggregates audio features at multiple scales much better than features using only a single window size.
% These results show that the model can well capture the machine dependencies of sound events at different scales.
Table.~\ref{ablation} shows the effects of AHAM and MWAM and their components on our model performance.
Obviously, temporal contexts can improve the performance of the model.
% In AHAM, the module only uses the entire question feature as a query \textit{(MWAFM}) to calculate attention score on audio better than cross-modal attention \textit{(w/ bi-queried)}.
In AHAM, the model only uses the entire question feature as a query \textit{(MWAFM}) to calculate the attention score on audio feature sequence, which outperforms their cross-modal attention \textit{(w/ bi-queried)}. At the same time, the performance is further degraded if both are not used \textit{(w/o un-queried)}.
It shows that it is tending to associate the semantics of the sentence with a particular segment of the audio.
In MWAM, we conducted a large number of experiments with multiple random seeds to take the average value, and calculated the value of the standard deviation under each evaluation index to be about 0.5.
It can be seen that the proposed model aggregating audio features at multiple scales is much better than using only features of a single window size.
Furthermore, the experiment demonstrates that the proposed MWAFM can leverage unimodal and cross-modal to capture the machine dependencies of sound events at different scales.
Additionally, we provide convincing results by using multiple large window sizes, like MWAFM$^{\prime}$ in Table.~\ref{ablation}. The result shows the combined effect of attention module networks with multi-scale window sizes is the best. The fact is that short-term sound events tend to be more related to adjacent sound segments and less to distant. Meanwhile, considering the long duration of some acoustic events, the proposed model retains the large window attention mechanism.
Therefore, an effective attentive multi-scale window attention module captures sound events of various lengths and their temporal contextual dependencies well.
Apart from this, we replace the MWAM with a conventional transformer that stacks asynchronous hybrid attention modules (MWAFM*), and set the stacked layers to 4 for fair comparison with 4 different window sizes in MWAM. 
We conduct experiments on Cloto-AQA and find that this way achieves \textit{Top}-1, \textit{Top}-5, and \textit{Top}-10 scores of 21.22\%, 46.08\%, and 56.44\%, respectively. Compared to using MWAM, the results are lower by 1.02\%, 0.48\%, and 1.31\%, primarily because of the variable length of events in the audio that make it challenging for fixed-size window attention stacking to capture them. 
Hence, it is necessary to use a multi-scale window attention module, which can enhance the model’s performance.

\begin{table}[t]
\label{music-avqa}
\begin{center}
\caption{AQA results of different methods on the test set of Audio-MUSIC-AVQA. The top-1 results are highlighted.}
\vspace{-0.6em}
\begin{tabular}{l|lll}
\hline
Method        & Counting & Comparative & Average \\ \hline
GRU~\cite{antol2015vqa}             & 65.00    & 64.31       & 64.74   \\
BiLSTM~\cite{hori2017attention}     & 65.59    & 45.62       & 58.22   \\
HCAttn~\cite{lu2016hierarchical}    & 66.08    & 56.23       & 62.45   \\
HME~\cite{fan2019heterogeneous}     & 67.65    & 63.97       & 66.29   \\
PSAC~\cite{li2019beyond}	        & 69.03	   & 60.94       & 66.05   \\
FCNLSTM~\cite{fayek2020temporal}    & 66.96    & 62.96       & 65.49   \\
CONVLSTM~\cite{fayek2020temporal}   & 68.04    & 62.63       & 66.05   \\
ST-AVQA~\cite{Li_2022_CVPR}         & 67.75    & \textbf{64.65}       & 66.60   \\
AquaNet~\cite{lipping2022clotho}    & 65.59    & 52.86       & 60.89   \\ 
\hline
\textbf{MWAFM} & \textbf{69.42}    & 64.31       & \textbf{67.54}   \\ \hline
\end{tabular}
\end{center}
\vspace{-1.75em}
\end{table}

% \textcolor{blue}{
% \textbf{More convincing results}: 
% % In Table 2, the performances of WS=6 look strange. 
% % For WS=2,4,12, the \textit{Top}-10 is almost the same (< 0.2\% difference), while for WS=6, the \textit{Top}-10 dramatically drops more than 1\%. 
% % Such a large difference is even larger than the difference between WS=12 and WS=[2,4,6,12]. 
% % The authors should provide more convincing results, e.g., reporting the average performance of multiple runs as well as the standard deviation.
% Thanks for the reviewer’s thoughtful advice.
% % We provide the average performance and standard deviation of the results of multiple experimental runs to provide more convincing results.
% We conducted experiments with multi random seeds (\textit{ws=4/6/12)}, 
% and the average performance (21.86/21.18/21.32, 45.42/44.59/45.08, 56.11/55.80/55.99) 
% and standard deviation (0.77/0.23/0.42, 0.71/0.61/0.57, 0.44/
% 0.55/0.76) 
% to provide more convincing results.
% }

\textbf{Generalization of the model.}
% We expect the proposed method to perform well on other real-voice-based AQA tasks, so it is validated on the Audio-MUSIC-AVQA dataset. 
We expect the proposed method to be effective in other similar real-world sound scene understanding. Audio-MUSIC-AVQA proposed by Li et al.~\cite{Li_2022_CVPR} contains rich dynamic and complex audio scenes, which is very suitable for verifying the performance of MWAFM.
Therefore, we conduct extensive experiments on this dataset.
As shown in Tabel. 3, the overall results of the proposed MWAFM method achieve the best.  
Although the performance on some sub-problems is not the best, it also achieves satisfactory results.
They are sufficient to illustrate the generalization performance of the proposed MWAFM method.

% \input{tables/rebuttal_r2}
% \input{tables/rebuttal_r2_s6}
% \input{tables/rebuttal_r2_flops}
% \input{tables/rebuttal_r2_flops_v1}

% \textbf{Computational costs}: 
% We provide the computational cost (FLOPs and params) of the proposed model.
% The MWAFM's FLOPs and params are 0.0674\textit{G} and 3.1959\textit{M}, respectively.
% And the MWAFM w/o multi-scale window attention moudle's FLOPs and params are 0.02955\textit{G} and 1.6179\textit{M}, respectively.
% Although the computational cost increases, the overall is not high.
% Hence, the proposed MWAFM achieves excellent performance with minimal extra computation cost.

\section{Conclusion}
% \vspace{-0.55em}
\label{sec:Conclusion}
In this paper, we present a novetly \textbf{M}ulti-scale \textbf{W}indow \textbf{A}ttention \textbf{F}usion \textbf{M}odel (\textbf{MWAFM}) for audio question answering task.
The proposed MWAFM provides a powerful ability for AQA tasks, which can capture sound events of various lengths and their temporal contextual dependencies well and assign large weights to key audio snippets that are more relevant to the question.
We evaluate the MWAFM on two large-scale audio datasets (Clotho-AQA and Audio-MUSIC-AQA), which illustrates the effectiveness and generalization of the proposed model.
We believe our research will facilitate the development of fine-grained audio scene understanding, especially in terms of temporal reasoning.

\noindent \textbf{Acknowledgement.}
This work is supported by the Fundamental Research Funds for the Central Universities, and the Research Funds of Renmin University of China (NO. 22XNH031).

% \textcolor{blue}{
% To validate the effectiveness of the proposed model, we conducted a large number of experiments, including comparison with mainstream QA methods, various ablation study module et al,. 
% Sufficient experimental results demonstrate the effectiveness and generalization of the proposed MWAFM.
% Simultaneously, the third contribution in the \textit{Introduction section} is also expressed with reference to the description of relevant works.
% }

\bibliographystyle{IEEEtran}
\bibliography{mybib}

\end{document}